\documentclass[prd,draft,preprint,showpacs,groupedaddress]{revtex4-1}
\usepackage{amsmath}
\usepackage{amsfonts}
\usepackage{amssymb}
\usepackage{geometry}
\usepackage{natbib}
\usepackage[english]{babel}
\begin{document}

  \title{Conformal transformation route to gravity's rainbow}

  \author{Miao He} \author{Ping Li} \author{Zi-Liang Wang} \author{Jia-Cheng Ding}
  \author{Jian-Bo Deng} \email[Jian-Bo Deng: ]{dengjb@lzu.edu.cn}
  
  \affiliation{Institute of Theoretical Physics, LanZhou University, Lanzhou 730000, P. R. China}

  \date{\today}

  \begin{abstract}
  Conformal transformation as a mathematical tool has been used in many areas of gravitational 
  physics. In this paper, we would consider the gravity's rainbow, in which the metric could be 
  treated as a conformal rescaling of the original metric. By using the 
  conformal transformation technique, we get a specific form of modified Newton's constant and 
  cosmological constant in gravity's rainbow, which implies that the total vacuum energy is 
  dependent on probe energy. Moreover, the result shows that the Einstein gravity's rainbow 
  could be described by an energy-dependent $f(E,\tilde R)$ gravity. At last, we study the 
  $f(R)$ gravity, when the gravity's rainbow is considered, it can also be described as another 
  energy-dependent $\tilde f(E,\tilde R)$ gravity.   
  \end{abstract}
  
  

  \maketitle

  \section{Introduction}
  
  The classical general relativity has been proved to describe our low energy world quite well. 
  However, recent astronomical and cosmological observations, such as the threshold anomalies 
  of ultrahigh energy cosmic rays and TeV photons~\cite{14,15,16,17,18,19}, would cause some 
  puzzles in general relativity. To construct a semi-classical or an effective theory of quantum 
  gravity where the Planck length might play a fundamental role without violating the principle 
  of relativity, 
  deformed or doubly special relativity was proposed~\cite{1,2,20,21}. This theory leads to a 
  deformed Lorentz symmetry such that the usual energy-momentum relation or dispersion relation 
  in special relativity may be modified with corrections in the order of Planck length.
  \par
  The deformed special relativity formalism was generalized to curved spacetime by Magueijo 
  and Smolin, whose formalism was called gravity's rainbow~\cite{4}. It shows that there will 
  be an energy-dependent metric. Furthermore, these lead to an energy-dependent connection, 
  curvature and a modification to the Einstein equations. The vacuum energy depending on 
  modified dispersion relations at high energy predicted by gravity's rainbow was discussed in
  ~\cite{40}.
  As a modified gravity theory, it has been considered in many areas and got many interesting 
  results. Rainbow deformation of various black hole solutions have been performed~\cite{9}. 
  In gravity's rainbow, the uncertainty principle still holds~\cite{35,36}, which can 
  transform the test particle energy into the radius of the event horizon. 
  This uncertainty principle relation has been used in the black hole thermodynamics
  ~\cite{23,26,45} and black hole remnants~\cite{27,37}. Gravity's rainbow has been also 
  considered to study the early universe, such as the nonsingular universes~\cite{8,22,33,34,42} 
  or big bounce universes~\cite{13}. 
  Furthermore, the gravity's rainbow was investigated in Gauss-Bonnet gravity~\cite{24,25,33}, 
  massive gravity~\cite{38,39} and $f(R)$ gravity~\cite{28}. Recent research shows that the 
  gravity's rainbow has some   
  connections with other gravity theories and quantum gravity. A connection between $f(R)$ 
  gravity and gravity's rainbow has been firstly discussed in~\cite{41,42}. Also, there is a 
  correspondence of gravity's rainbow with Ho\v{r}ava-Lifshitz gravity~\cite{43}.
  \par
  Conformal rescalings and conformal techniques as a mathematical tool have been widely used in 
  general relativity~\cite{5}, especially in the scalar-tensor theory of gravity. It often maps 
  the equations of motion of physical systems into mathematically equivalent sets of equations 
  that are easier to solve or more convenient to study. This situation emerges mainly in 
  alternative theories of gravity, unified theories in multidimensional spaces. 
  By applying a conformal transformation, 
  problems would move from one conformal frame to another. The Jordan frame and the Einstein 
  frame are those discussed most frequently among many conformal frames. 
  The conformal rescaling to the nonminimal coupling case 
  for the scaler field in Brans-Dicke theory can get the minimal coupling case of the scaler 
  field~\cite{29,46}, but the the scaler field may couple to the matter field.
  Taking into account conformal transformation of Brans-Dicke theory with an electrodynamics 
  Lagrangian, scalar field should couple with electrodynamics in dilaton gravity, which has been 
  discussed in~\cite{44}.
  Physical equivalence between nonlinear gravity and a general-relativistic self-gravitating 
  scalar field was proved by conformal technique as well~\cite{6}. 
 
  \par
  In this paper, we would use the conformal transformation technique to investigate the 
  gravity's rainbow, in which the metric is a conformal rescaling of the 
  original one. Through the conformal transformation, we get a specific form of modified 
  Newton's 
  constant and cosmological constant in gravity's rainbow, which implies an energy-dependent 
  vacuum energy. Furthermore, our result shows that the Einstein gravity's rainbow could be 
  described as a modified $f(E,\tilde R)$ gravity, which is energy-dependent; thus, a 
  connection between gravity's rainbow and $f(R)$ gravity was established. We also get the 
  Friedmann equations in Einstein gravity's rainbow under the modified $f(E,\tilde R)$ gravity 
  framework. Motived by this, 
  $f(R)$ gravity was considered and the $f(R)$ gravity's rainbow can be also treated as an 
  energy-dependent $\tilde f(E,\tilde R)$ gravity.       
  \par
  This paper is organized as follows: Sec.\uppercase\expandafter{\romannumeral2} 
  is a review of the gravity's rainbow.
  In Sec.\uppercase\expandafter{\romannumeral3}, The Einstein gravity's rainbow is investigated
  by using the conformal transformation. We also consider the Friedmann equation under 
  the modified $f(E,\tilde R)$ gravity framework.
  In Sec.\uppercase\expandafter{\romannumeral4}, we generalize this framework to the $f(R)$ 
  gravity's rainbow. Conclusions and discussions are given in Sec.\uppercase\expandafter
  {\romannumeral5}.

  \section{Review of gravity's rainbow}
  
  Deformed or doubly special relativity is a theory that implies a modified set of 
  principles of special relativity~\cite{1,2}. As a result, the invariant of energy and  
  momentum in general may be modified to
  \begin{equation}
    \label{eq:invariant of energy and momentum}
    E^2f_{1}^{2}(E)-p^2f_{2}^{2}(E)=m_{0}^2 \qquad,
  \end{equation}
  where the two general functions $f_{1}^{2}(E)$ and $f_{2}^{2}(E)$ depend on the energy of 
  probes $E$. The correspondence principle requires that $f_{1}^{2}(E),f_{2}^{2}(E) \to 1$ as
   $E\ll1$ with the Planck   
  scale $E_{Pl}=1$. To make sure the contraction between infinitesimal displacement and momentum
  is a linear invariant~\cite{3,4}
  \begin{equation}
    \label{eq:invariant of displacement and momentum}
    dx^{\mu}p_{\mu}=dtE+dx^{i}p_{i} \qquad,
  \end{equation}
  the usual flat metric should be replaced by the rainbow metric defined as 
  \begin{equation}
    \label{eq:flat rainbow metric}
    ds^2=-\frac{1}{f_{1}^2(E)}dt^2+\frac{1}{f_{2}^2(E)}dx^2 \qquad.
  \end{equation}
  The flat rainbow metric indicates that the geometry of spacetime depends on the energy of 
  probes. That is to say, the geometry of spacetime probed by a particle with energy $E$ can 
  be described by an energy-dependent orthonormal frame fields 
  \begin{equation}
    \label{eq:orthonormal frames fields}
    \tilde e_{0}(E)=\frac{1}{f_{1}(E)}e_{0},\qquad \tilde e_{i}(E)=\frac{1}{f_{2}(E)} e_{i}  
    \qquad,
  \end{equation}
  where $e_{0}$ and $e_{i}$ represent the energy-independent frame fields. 
  Therefore,
  the flat rainbow metric can be written as
  \begin{equation}
    \label{eq: rainbow metric energy-dependent frame field}
    g(E)=\eta^{\mu\nu}\tilde e_{\mu}(E) \otimes \tilde e_{\nu}(E)\qquad.
  \end{equation}
  \par
  The rainbow metric formalism can be generalized when the gravity is taken into account. It may 
  lead to the fact that connection $\nabla_{\mu}(E)$ and curvature tensor 
  $R^{\sigma}_{\mu\nu\lambda}(E)$ 
  are all energy-dependent. One can also define an energy-dependent energy-momentum 
  tensor $T_{\mu\nu}(E)$. Then the Einstein equations should be replaced by a one parameter 
  family of equations 
  \begin{equation}
    \label{eq: rainbow Einstein equations}
    G_{\mu\nu}(E)=8\pi G(E)T_{\mu\nu}(E)+g_{\mu\nu}(E)\Lambda(E) \qquad,
  \end{equation} 
  where the Newton's constant $G(E)$ and cosmological constant $\Lambda(E)$ is conjectured to be 
  energy-dependent. Futhermore, these are assumed as $G(E)=h_{1}^{2}(E)G$ and $\Lambda(E)=h_{2}^
  {2}(E)\Lambda$ from the view point of renormalization group theory~\cite{4,8}.
  
  \section{Einstein gravity's rainbow}
  
  In this section we would consider the relation between Einstein gravity and  
  Einstein gravity's rainbow. For simplicity, we start with the static spherically symmetric 
  metric. We may write the rainbow metric in either energy-dependent coordinates or 
  energy-independent
  coordinates. Generally, in energy-dependent coordinates the static spherically symmetric 
  metric takes the form
  \begin{equation}
    \label{eq: rainbow metric in energy-dependent coordinates}
    d\tilde s^2=-A(\tilde r(E),E)d\tilde t(E)^2+B(\tilde r(E),E)d\tilde r(E)^2+\tilde r(E)^2d
    \Omega^2 \qquad.
  \end{equation}  
  In terms of energy independent coordinates the general form for a static spherically symmetric 
  metric is~\cite{4} 
  \begin{equation}
    \label{eq: rainbow metric in energy independent coordinates}
    d\tilde s^2=-\frac{A(r)}{f_{1}^2(E)}dt^2+\frac{B(r)}{f_{2}^2(E)}dr^2+\frac{r^2}{f_{2}^2(E)}d
    \Omega^2 \qquad.
  \end{equation}
  In the limit of $E\ll1$, this rainbow metric would reduce to 
  \begin{equation}
    \label{eq: original metric}
    ds^2=-A(r)dt^2+B(r)dr^2+r^2d\Omega^2 \qquad,
  \end{equation}
  which is the case in Einstein gravity.  
  \par
  Conformal rescalings and conformal techniques as a mathematical tool have been widely used in 
  general relativity for a long time~\cite{5}. It often maps the equations of motion of physical 
  systems into mathematically equivalent sets of equations that are more easily solved or 
  more convenient to study. A well-known example is that the nonlinear gravity theories in 
  Jordan frame could be equivalent to the Einstein frame with a scalar field under the conformal 
  transformation~\cite{6,7}. For the gravity's rainbow, an interesting case of the rainbow 
  function is~\cite{1}
  \begin{equation}
    \label{eq: case of dispersion relations}
    f_{1}=f_{2}=\frac{1}{1+\lambda E} \qquad,
  \end{equation}
  which would not produce a varying speed of light $c$. In this case, the rainbow metric is 
  just a conformal rescaling by an energy-dependent function
  \begin{equation}
    \label{eq: conformal rescale}
    d\tilde s^2=\frac{1}{f_{2}^2(E)}ds^2 \qquad.
  \end{equation}
  \par
  To consider the Einstein gravity's rainbow, let us start with 
  the Einstein-Hilbert action
  \begin{equation}
    \label{eq: E-H action}
    S=\frac{1}{16\pi G}\int\sqrt{-g}(R+2\Lambda)d^4x+S_{M}\qquad,
  \end{equation}
  where $\Lambda$ is the cosmological constant and $S_{M}$ represents the matter field action. 
  It is possible to derive an action in gravity's 
  rainbow framework under the conformal transformation
  \begin{equation}
    \label{eq: conformal transformation}
    \tilde g_{\mu\nu}=f_{2}^{-2}(E)g_{\mu\nu} \qquad,
  \end{equation}
  where $\tilde g_{\mu\nu}$ represents the rainbow metric and 
  $g_{\mu\nu}$ is the original metric.
  As $f_{2}(E)$ is independent of the coordinates $\{t, r, \theta, \phi\}$, one can get the 
  following relations
  \begin{eqnarray}
    \label{eq: conformal transformation relations}
    R=f_{2}^{-2}(E)\tilde R,\qquad \sqrt{-g}=f_{2}^{4}(E)\sqrt{-\tilde g} \qquad.
  \end{eqnarray}
  Then one can get the action in gravity's rainbow framework
  \begin{equation}
    \label{eq: E-H action in rainbow frame}
    \tilde S=\frac{1}{16\pi G}\int\sqrt{-\tilde g}f_{2}^{4}(E)(f_{2}^{-2}(E)\tilde R+2
    \Lambda)d^4x
    +\tilde S_{M} \qquad.
  \end{equation}
  Varying this action with respect to $\tilde g_{\mu\nu}$, one obtains 
  \begin{equation}
    \label{eq: equation of motion in rainbow frame}
    \tilde R_{\mu\nu}(E)-\frac{1}{2}\tilde R(E)\tilde g_{\mu\nu}=f_{2}^{2}(E)\Lambda\tilde 
    g_{\mu\nu}+8\pi G f_{2}^{-2}(E)\tilde T_{\mu\nu} \qquad.
  \end{equation}
  This equation would be consistent with eq.\eqref{eq: rainbow Einstein equations}, if we set 
  \begin{eqnarray}
    \label{eq: conformal transformation relations}
    G(E)=G f_{2}^{-2}(E),\qquad \Lambda(E)=\Lambda f_{2}^{2}(E)\qquad.
  \end{eqnarray}
  This result agrees with the previous assumption about $G(E)$ and $\Lambda(E)$~\cite{4,8}.
  \par
  Moreover, the energy-dependent Newton's constant and cosmological constant would imply an 
  energy-dependent total vacuum energy $\tilde E_{vac}(E)$ rather than a constant one~\cite{9}. 
  The vacuum energy including a dependence on gravity's rainbow has been discussed in~\cite{40}.
  In fact, giving a rainbow metric as 
  eq.\eqref{eq: rainbow metric in energy independent coordinates}, we note that in general the 
  volume of a spatial region with fixed size $r$ is energy-dependent as 
  $\tilde V\sim f_{2}^{-3}(E)V$. Then the total vacuum energy should be 
  \begin{equation}
    \label{eq: total vacuum energy}
     \tilde E_{vac}\sim \tilde\rho_{vac}\tilde V\sim\frac{\Lambda(E)}{8\pi G(E)}f_{2}^{-3}(E)V
     \sim f_{2}(E)E_{vac}\qquad.
  \end{equation}
  \par
  In parallel, we would like to consider the gravity's rainbow from another side. 
  As the modified 
  Einstein equations eq.\eqref{eq: equation of motion in rainbow frame} could be obtained by 
  varying the action of Einstein gravity's rainbow eq.\eqref{eq: E-H action in rainbow frame}. 
  The Einstein gravity's rainbow could be 
  treated as a modified $f(E,\tilde R)$ gravity which is energy-dependent, if we set 
  $f(E,\tilde R)=f_{2}^{4}(E)(f_{2}^{-2}(E)\tilde R+2\Lambda)$. In the model of $f(R)$ gravity, 
  the field  equation could be written as~\cite{10,11}
  \begin{equation}
    \label{eq: f_r field equation}
     R_{\mu\nu}-\frac{1}{2}g_{\mu\nu}R=\frac{1}{f_{R}}\left(T_{\mu\nu}^{curv}+8\pi G T_{\mu\nu}
     \right)\qquad,
  \end{equation}
  where $f_{R}=df/dR$ and $T_{\mu\nu}^{curv}$ is curvature energy-momentum tensor defined as
  \begin{equation}
    \label{eq: curvature energy-momentum tensor}
     T_{\mu\nu}^{curv}=\frac{1}{2}g_{\mu\nu}(f-Rf_{R})+(g_{\mu\rho}g_{\nu\sigma}-g_{\mu\nu}
     g_{\rho\sigma})\nabla^{\rho}\nabla^{\sigma}f_{R} \qquad.
  \end{equation} 
  Similarly, in the Einstein gravity's rainbow, one can get $f_{\tilde R}=f_{2}^{2}(E)$ and 
  $\tilde T_{\mu\nu}^{curv}=\Lambda f_{2}^{4}(E)\tilde g_{\mu\nu}$. Then one can get the field 
  equation
  \begin{equation}
    \label{eq: f_er rainbow field equation}
     \tilde R_{\mu\nu}-\frac{1}{2}\tilde g_{\mu\nu}\tilde R=\frac{1}{f_{2}^{2}(E)}\left(
     \Lambda f_{2}^{4}(E)\tilde g_{\mu\nu}+8\pi G \tilde T_{\mu\nu}
     \right)\qquad,
  \end{equation}
  which is the same as eq.\eqref{eq: equation of motion in rainbow frame}. However, there is no 
  need to introduce an energy-dependent Newton's constant and cosmological constant in this 
  framework. 
  \par
  Therefore, in the case of $f_{1}(E)=f_{2}(E)$, the Einstein gravity's rainbow is just a 
  conformal rescale for a static spherically symmetric metric, and it can be described by an 
  energy-dependent $f(E,R)$ gravity. In addition, we will show that this result also holds for 
  the case of $f_{1}(E)\ne f_{2}(E)$.       
  \par
  Considering the static spherically symmetric metric eq.\eqref{eq: rainbow metric in energy 
  independent coordinates}, if we introduce an energy-dependent time coordinate
  \begin{equation}
    \label{eq: energy-dependent time coordinate}
     \hat t(E) = \frac{f_{2}(E)}{f_{1}(E)}t\qquad,
  \end{equation}
  eq.\eqref{eq: rainbow metric in energy 
  independent coordinates} could be written as 
  \begin{equation}
    \label{eq: energy-dependent time coordinate}
    d\tilde s^2=\frac{1}{f_{2}^2(E)}\left[-A(r)d\hat t^2+B(r)dr^2+r^2
    d\Omega^2\right]=\frac{1}{f_{2}^2(E)}d\hat s^2 \qquad.
  \end{equation}
  It is also a conformal transformation  of the original metric with the coordinates 
  $\{\hat t, r, \theta, \phi\}$. we should point out that the energy-dependent time $\hat t$ 
  won't change the time-like killing vector, as $f_{2}(E)/f_{1}(E)$ is independent with 
  coordinates~\cite{4}. Then the line element $d\hat s^2$ also satisfies 
  the Einstein equations.
  In fact, when $f_{1}(E)\ne f_{2}(E)$, the speed of light $c(E)=f_{2}(E)/f_{1}(E)$ is 
  energy-dependent~\cite{12}. It is naturally to introduce $\hat t=c(E)t$. 
  Once $f_{1}(E)=f_{2}(E)$, we have 
  $\hat t=t$, which would reduce to the previous case. Therefore, the above discussion also 
  holds for $f_{1}(E)\ne f_{2}(E)$, and the price is to introduce an energy-dependent time 
  coordinate $\hat t$. We would like to point out that the original time coordinate should be 
  substituted back after all the calculations, especially for the cosmological time.   
  \par
  As an example, we consider the rainbow universe under modified $f(E,R)$ gravity framework. 
  The modified flat FRW metric for the gravity's rainbow could be expressed as~\cite{4,8,13}
  \begin{equation}
    \label{eq: modified flat FRW metric}
    d\tilde s^2=\frac{1}{f_{2}^2(E)}\left[-d\hat t^2+a^{2}(\hat t)(dr^2+r^2
    d\Omega^2)\right]\qquad.
  \end{equation}
  With the substitution $\tilde t= f_{2}^{-1}(E)\hat t$ and $\tilde a(\tilde t)=f_{2}^{-1}(E)a(\hat t)$,  
  the modified flat FRW metric is just
  \begin{equation}
    \label{eq: modified flat FRW metric_m}
    d\tilde s^2=-d\tilde t^2+\tilde a^2(\tilde t)(dr^2+r^2d\Omega^2)\qquad,
  \end{equation}
  which should satisfy the Friedmann equations in a modified $f(E,\tilde R)$ gravity.
  Generally, the Friedmann equations for $f(R)$ gravity are~\cite{7}
  \begin{equation}
    \label{eq: f_r Friedmann equation_1}
    H^2=\frac{Rf_{R}-f}{6f_{R}}-H\frac{\dot{f_{R}}}{f_{R}}+\frac{8\pi G}{3f_{R}}\rho \qquad,
  \end{equation}
  \begin{equation}
    \label{eq: f_r Friedmann equation_2}
    \dot{H}=-\frac{\ddot{f_{R}}-H\dot{f_{R}}}{2f_{R}}+\frac{H\dot{f_{R}}}{2f_{R}}-\frac{4\pi G}{f_{R}}
    (\rho+p) \qquad.
  \end{equation}
  For the Einstein gravity's rainbow, we should set $f(E,\tilde R)=f_{2}^{4}(E)(f_{2}^{-2}(E)
  \tilde R+2\Lambda)$. The energy-momentum tensor has a perfect fluid form
   \begin{equation}
    \label{eq: rainbow energy-momentum tensor}
    \tilde T_{\mu\nu}=\tilde\rho u_{\mu} u_{\nu}+\tilde p(\tilde g_{\mu\nu}+u_{\mu}u_{\nu})
    \qquad,
  \end{equation} 
  where $u_{\mu}$ depends on $E$ and is defined as $u_{\mu}=(f_{2}^{-1}(E), 0, 0, 0)$ with the 
  time coordinate $\hat t$, such 
  that $\tilde g^{\mu\nu}u_{\mu}u_{\nu}=-1$~\cite{4}.
  Thus, the modified Friedmann equations could be
  \begin{equation}
    \label{eq: rainbow Friedmann equation_1}
    \left(\frac{\dot{\tilde a}}{\tilde a}\right)^2=\frac{f_{2}^2(E)\Lambda}{3}+\frac{8\pi G}
    {3f_{2}^2(E)}\tilde\rho \qquad,
  \end{equation}
  \begin{equation}
    \label{eq: rainbow Friedmann equation_2}	
    \frac{\ddot{\tilde a}}{\tilde a}-\left(\frac{\dot{\tilde a}}{\tilde a}\right)^2=
    -\frac{4\pi G}
    {f_{2}^2(E)}(\tilde\rho+\tilde p)\qquad.
  \end{equation}
  These would produce a conservation equation
  \begin{equation}
    \label{eq:  conservation equation of energy-momentum tensor}
    \dot{\tilde \rho}= -3\frac{\dot {\tilde a}}{\tilde a}(\tilde\rho+\tilde p)
    \qquad.
  \end{equation} 
  If we substitute back to the cosmological time and scale factor $a$, one can get the Friedmann equations
  \begin{equation}
    \label{eq: rainbow Friedmann equation_t_1}
    \left(\frac{\dot{a}}{a}\right)^2=\frac{f_{2}^2(E)\Lambda}{3f_{1}^{2}(E)}+\frac{8\pi G}
    {3f_{1}^{2}(E)f_{2}^2(E)}\tilde\rho \qquad,
  \end{equation}
  \begin{equation}
    \label{eq: rainbow Friedmann equation_t_2}	
    \frac{\ddot{a}}{a}-\left(\frac{\dot{a}}{a}\right)^2=
    -\frac{4\pi G}
    {f_{1}^{2}(E)f_{2}^2(E)}(\tilde\rho+\tilde p)\qquad,
  \end{equation}
  which are consistent with the Friedmann equations derived by other methods~\cite{4,8,13}.
  They could get some interesting results with some specific rainbow functions, like no 
  singularity cosmological solution~\cite{8} and big bounce universe~\cite{13}.   
  \par
  Thus, the Einstein gravity's rainbow could be treated as an energy-dependent 
  $f(E,\tilde R)=f_{2}^{4}(E)(f_{2}^{-2}(E)\tilde R+2\Lambda)$ gravity model through the 
  conformal transformation technique. This may give a convenient route to study the gravity's 
  rainbow.
      
  \section{$f(R)$ Gravity's rainbow}
  
  For more general cases, we consider the $f(R)$ gravity's rainbow. Let us start with the action
  in $f(R)$ gravity
  \begin{equation}
    \label{eq: f_r action}
    S=\frac{1}{16\pi G}\int\sqrt{-g}f(R)d^4x+S_{M}\qquad,
  \end{equation}
  where $S_M$ is the action of matter fields.
  The field equation can be derived by varying the action with respect to $g_{\mu\nu}$  
  \begin{equation}
    \label{eq: f_r field equation}
     R_{\mu\nu}-\frac{1}{2}g_{\mu\nu}R=\frac{1}{f_{R}}\left(T_{\mu\nu}^{curv}+8\pi G T_{\mu\nu}
     \right)\qquad,
  \end{equation}
  where $T_{\mu\nu}^{curv}$ is curvature energy-momentum tensor defined as eq.\eqref{eq: curvature energy-momentum tensor}.
  When the gravity's rainbow is considered, it may lead to a conformal transformation
  $\tilde g_{\mu\nu}=f_{2}^{-2}g_{\mu\nu}$ like the Einstein gravity's rainbow. Noting the 
  relation eq.\eqref{eq: conformal transformation relations}, then the action becomes
  \begin{equation}
    \label{eq: f_r rainbow action}
    \tilde S=\frac{1}{16\pi G}\int\sqrt{-\tilde g}f_{2}^{4}(E)f(f_{2}^{-2}(E)\tilde R)d^4x
    +\tilde S_{M}\qquad.
  \end{equation}
  Varying this action with respect to $\tilde g_{\mu\nu}$, one obtains
  \begin{equation}
    \label{eq: f_r rainbow field equation}
    \tilde R_{\mu\nu}-\frac{1}{2}\tilde R \tilde g_{\mu\nu}=\frac{1}{f_{2}^{2}(E)f_{R}(f_{2}^{-2}
    (E)\tilde R)}\left(\tilde T_{\mu\nu}^{curv}+8\pi G \tilde T_{\mu\nu}\right) \qquad,
  \end{equation}
  where
  \begin{eqnarray}
    \label{eq: f_r rainbow curvature energy-momentum tensor}
    \tilde T_{\mu\nu}^{curv}=\frac{1}{2}&f_{2}^{2}(E)\tilde g_{\mu\nu}[f_{2}^{2}(E)f(f_{2}^{-2}
    (E)\tilde R)-\tilde Rf_{R}(f_{2}^{-2}(E)\tilde R)]\nonumber \\
    +&f_{2}^{2}(E)(\tilde g_{\mu\rho}\tilde g_{\nu\sigma}-\tilde g_{\mu\nu}\tilde g_{\rho\sigma})
    \nabla^{\rho}\nabla^{\sigma}f_{R}(f_{2}^{-2}(E)\tilde R)\qquad.
  \end{eqnarray}
  It ends up with an energy-dependent modified $\tilde f(E,\tilde R)$ gravity. 
  Obviously, if $f(R)=R+2\Lambda$, it reduces to the Einstein gravity's rainbow, and the 
  curvature energy-momentum tensor $\tilde T_{\mu\nu}^{curv}$ would lead to 
  $\Lambda(E)=f_{2}^{2}(E)\Lambda$. 
  \par
  Thus, from the view point of action, the $f(R)$ gravity's rainbow can be treated as another 
  energy-dependent $\tilde f(E,\tilde R)$ 
  gravity, where $\tilde f(E,\tilde R)=f_{2}^{4}(E)f(f_{2}^{-2}(E)\tilde R)$. We should point out that 
  the time coordinate should be $\hat t =f_{2}(E)/f_{1}(E)t$ for the case of $f_{1}(E)\ne f_{2}(E)$. So 
  when a $f(R)$ gravity's rainbow was considered, it is convenient to turn to a corresponding 
  $\tilde f(E,\tilde R)$ gravity. This would open up a door to study the $f(R)$ gravity's rainbow.  
  As an interesting subject, $f(R)$ theories have been applied to dark energy~\cite{30,31}. 
  In this way, when the 
  $\tilde f(E,\tilde R)$ gravity is considered, it may lead to an energy-dependent dark energy which might 
  explain the late-time cosmic accleration~\cite{32}. 
  \section{Conclusions and discussions}
  
  In summary, we investigated the gravity's rainbow which was proposed by
  Magueijo and Smolin, and found that the rainbow metric is just a conformal rescaling of the 
  the original one for the case of $f_{1}(E)=f_{2}(E)$. Motivated by this, we introduced the 
  conformal transformation technique to study the gravity's rainbow. For the Einstein gravity's 
  rainbow, by using the conformal 
  transformation technique, we give the action of Einstein gravity's 
  rainbow. Varying this action, we got a specific form of modified Newton's constant and 
  cosmological constant which agrees with the renormalization group theory. 
  And these would imply an energy-dependent total vacuum energy rather than a constant one. 
  In addition, for the case of $f_{1}(E)\ne f_{2}(E)$, 
  there also exists a conformal transformation with an energy-dependent time 
  coordinate $\hat t=f_{2}(E)/f_{1}(E) t$, and the above results also hold. 
  \par
  From the view point of action, we found that the Einstein gravity's rainbow could be described 
  by an energy-dependent $f(E,\tilde R)$ gravity, where $f(E,\tilde R)=f_{2}^{4}(E)(f_{2}^{-2}(E)
  \tilde R+2\Lambda)$. The $f(E,\tilde R)$ gravity might provide an 
  effective route to research the gravity's rainbow. 
  A connection between gravity's rainbow and $f(R)$ theory was established.
  Moreover, under $f(E,\tilde R)$ gravity framework, we investigated the dynamic equations of 
  cosmology. For the FRW cosmology, the same Friedmann equations were obtained and these 
  equations could get some interesting cosmology solutions.
  \par
  For a more general case, we considered the $f(R)$ gravity's rainbow and found that it can 
  also be described by an energy-dependent $\tilde f(E,\tilde R)$ gravity where 
  $\tilde f(E,\tilde R)=f_{2}^{4}(E)f(f_{2}^{-2}(E)\tilde R)$. 
  When $f(R)=R+2\Lambda$, it can reduce to the Einstein gravity's rainbow and the curvature 
  energy-momentum tensor of $f(E,\tilde R)$ leads to the modified cosmological constant. As 
  many dark energy models were presented in $f(R)$ gravity, when the gravity's rainbow is 
  considered, the dark energy or vacuum energy might depend on the energy of probe~\cite{40}. 
  These would be left for further research in our later works.
       
  \section*{ACKNOWLEDGMENTS}
  
  We would like to thank the National Natural Science Foundation of
  China~(Grant No.11571342) for supporting us on this work.

  \bibliographystyle{unsrt}
  \bibliography{reference}

\end{document}